\documentstyle[preprint,aps,floats,psfig]{revtex}
\catcode`@=11
\def\references{%
\ifpreprintsty
\bigskip\bigskip
\hbox to\hsize{\hss\large \refname\hss}%
\else
\vskip24pt
\hrule width\hsize\relax
\vskip 1.6cm
\fi
\list{\@biblabel{\arabic{enumiv}}}%
{\labelwidth\WidestRefLabelThusFar  \labelsep4pt %
\leftmargin\labelwidth %
\advance\leftmargin\labelsep %
\ifdim\baselinestretch pt>1 pt %
\parsep  4pt\relax %
\else %
\parsep  0pt\relax %
\fi
\itemsep\parsep %
\usecounter{enumiv}%
\let\p@enumiv\@empty
\def\theenumiv{\arabic{enumiv}}%
}%
\let\newblock\relax %
\sloppy\clubpenalty4000\widowpenalty4000
\sfcode`\.=1000\relax
\ifpreprintsty\else\small\fi
}
\catcode`@=12
\begin{document}
\def\mh{m_h^{}}
\def\vev#1{{\langle#1\rangle}}
\def\gev{{\rm GeV}}
\def\tev{{\rm TeV}}
\def\fbi{\rm fb^{-1}}
\def\lsim{\mathrel{\raise.3ex\hbox{$<$\kern-.75em\lower1ex\hbox{$\sim$}}}}
\def\gsim{\mathrel{\raise.3ex\hbox{$>$\kern-.75em\lower1ex\hbox{$\sim$}}}}
\newcommand{\teb}{{T_{\bar{\nu}_e}}}
\newcommand{\neb}{{\bar{\nu}_e}}
\newcommand{\tneb}{{T_{\bar{\nu}_e}}}
\newcommand{\tnx}{{T_{\bar{\nu}_x}}}
\newcommand{\eneb}{{\vev{E_{\bar{\nu}_e}}}}

\newcommand{ \slashchar }[1]{\setbox0=\hbox{$#1$}   
   \dimen0=\wd0                                     
   \setbox1=\hbox{/} \dimen1=\wd1                   
   \ifdim\dimen0>\dimen1                            
      \rlap{\hbox to \dimen0{\hfil/\hfil}}          
      #1                                            
   \else                                            
      \rlap{\hbox to \dimen1{\hfil$#1$\hfil}}       
      /                                             
   \fi}                                             %

\tighten
\preprint{ \vbox{
\hbox{MADPH--02-1257}
\hbox{hep-ph/0202158}}}
\title{Supernova 1987A did not test the neutrino mass hierarchy}
\author{$^1$V. Barger, $^2$D. Marfatia and $^1$B. P. Wood}
\vskip 0.3in
\address{$^1$Department of Physics, University of Wisconsin, Madison, WI 53706}
\vskip 0.1in
\address{$^2$Department of Physics, Boston University, Boston, MA 02215}
\maketitle

\begin{abstract}
{\rm We dispel the misconception that data from SN 1987A favor the normal neutrino mass hierarchy 
over the inverted hierarchy for $\sin^2 \theta_{13} \gsim 10^{-4}$. 
We find comparable fits for the two hierarchies. 
No bound can be placed on the mixing angle $\theta_{13}$ even at the 1$\sigma$ level.
}
\end{abstract}

On February 23, 1987, antineutrinos from a 20$M_{\odot}$ Type II supernova in the Large Magellanic
Cloud were detected by the Kamiokande II (KII)~\cite{snkam} and Irvine Michigan Brookhaven (IMB)
~\cite{snimb} detectors.
The data on SN 1987A have been important in confirming the generic features of the 
core collapse model of supernovae~\cite{lamb}.
However, the average antineutrino energy $\eneb$ and the binding 
energy of the star $E_b$,
extracted from the signals at KII and IMB, only marginally agree with each 
other~\cite{raffelt,lunardini,valle}.
Moreover, even under the most optimistic conditions, the allowed values of $\eneb$
 are lower than the following predictions of traditional
supernova models~\cite{models1}:
\begin{eqnarray}
\eneb &=& 14-17\  {\rm MeV}\,,  \nonumber \\
\vev{E_{\bar{\nu}_x}} &=& 24-27 \  {\rm MeV}\,, \ \ \ \ \ \ \  x=\mu,\  \tau\,,  \nonumber \\
E_b &=& 1.5-4.5 \times 10^{53}\  {\rm ergs} \label{models} \,.
\end{eqnarray}
Including neutrino oscillations does little to 
resolve these discrepancies. For the favored large-angle solution to the solar anomaly~\cite{lma},
compatibility of the two data sets is not significantly improved~\cite{raffelt,lunardini,valle}. 
Furthermore, in any oscillation scenario,  
$\neb \leftrightarrow \bar{\nu}_{\mu,\tau}$ oscillations result in a
hardened initial spectrum, implying even lower $\eneb$ and 
increasing the discrepancy with theoretical 
predictions~\cite{raffelt,lunardini,valle}. 
The degree to which the data and supernova models disagree depends on the type of 
neutrino mass hierarchy. There are two possibilities depending upon whether $m_3$ is the
largest or smallest mass eigenstate:
\begin{eqnarray}
\Delta m^2_{32}\equiv m_3^2-m_2^2 &>&0\,,\ \ \ \ \ \ {\rm normal\ hierarchy,}\\
                            &<&0\,,\ \ \ \ \ \ {\rm inverted\ hierarchy,}
\end{eqnarray}
where oscillations between the $m_3$ and $m_2$ ($m_2$ and $m_1$) mass eigenstates 
are responsible for the
atmospheric (solar) neutrino deficit.
The disagreement with the theoretical predictions is aggravated
if the hierarchy of neutrino masses is 
inverted and $\sin^2 \theta_{13} \gsim 10^{-3}$ because
 the conversion of
the original $\neb$ spectrum is almost complete in this case (regardless of the solution to the
solar neutrino problem). This is a hint that the inverted mass scheme is disfavored by
SN 1987A~\cite{raffelt,lunardini}. However, it is
an overstatement that SN 1987A gives a strong indication that the
 inverted hierarchy is disfavored unless $\sin^2 \theta_{13} \lsim 10^{-4}$~\cite{minakata}. 

We show that the SN 1987A data do not favor one hierarchy over the other by performing a maximum
likelihood analysis of the data from KII and IMB in the three neutrino framework assuming
the LMA solution to the solar neutrino problem. We perform three-parameter fits to 
$\teb$, $E_b$ and $\sin^2 2 \theta_{13}$, where $\teb$ is the
temperature of the electron antineutrinos originating in the supernova, which for a Fermi-Dirac 
distribution is related to
$\eneb$ via $\teb=180\zeta (3)\eneb/(7 \pi^4) \approx \eneb/3.15$.
We find that the data can be fit equally well by the two hierarchies with the maximum of the 
likelihood function in the two cases to be approximately the same. No constraint 
can be placed on $\sin^2 2 \theta_{13}$ at the 1$\sigma$ level. 

Recent refinements in supernova codes suggest that the discrepancy with 
SN 1987A data may soon be resolved. 
The simulations of Ref.~\cite{models2} find $\eneb$ and
 $\vev{E_{\bar{\nu}_x}}$ below the ranges in Eq.~(\ref{models}). 
The lower $\tnx$ is attributed to the inclusion (for the first time) of 
nucleon-nucleon neutral-current bremsstrahlung which is softer than 
$e^+e^-$ annihilation (the other major $\bar{\nu}_x$ source).
A similar conclusion is drawn in Ref.~\cite{models3} which explains that nucleon recoils 
lower $\tnx$  substantially. Additional energy transfer due to neutrino-matter scattering processes
not included in supernova simulations 
will further soften the emergent spectra.

To simulate the time averaged spectrum of antineutrinos 
from SN 1987A, we employ a
Fermi-Dirac distribution with equipartition of energy between the neutrino flavors. 
As the antineutrinos travel
outward from their production point in the supernova, they encounter a density profile that falls 
like $1/r^3$~\cite{bethe}. 
For the LMA solution, antineutrinos can experience at most one resonance 
(at density $\rho \approx 10^3 - 10^4 \; {\rm g/cm}^3$ and
characterized by $\Delta m^2_{32}$ and $\sin^2 2\theta_{13}$), 
and only if the hierarchy is inverted. 
For the normal
and inverted hierarchies, the survival probability of electron antineutrinos is given by~\cite{dighe}
\begin{equation}
\bar{p} = \bar{P}_{1e}\,,
\label{norprob}
\end{equation} 
and
\begin{equation}
\bar{p}=P_H\bar{P}_{1e} + (1-P_H) \sin^2 \theta_{13}\,,
\label{invprob}
\end{equation}
respectively. Here, 
$\bar{P}_{1e} = \bar{P}_{1e}(E_{\nu},\Delta m^2_{21},\sin^2 2 \theta_{12})$ 
is the probability that an 
antineutrino reaching the earth in the $\bar{\nu}_1$ mass eigenstate will interact with the detector
as a $\neb$, and $P_H\sim e^{-\sin^2 2 \theta_{13}(|\Delta m^2_{32}|/E_\nu)^{2/3}}$~\cite{kuo} 
is the hopping
 probability at the resonance. 
If $\sin^2 2 \theta_{13} \ll 10^{-3}$, we have $P_H \approx 1$ and the survival probabilities 
for the two hierarchies are the same. Thus, the normal and inverted mass 
hierarchies are indistinguishable for $\sin^2 2 \theta_{13} \ll 10^{-3}$. 
If $\sin^2 2 \theta_{13} \gsim 10^{-3}$, 
for the inverted hierarchy $\bar{p} \approx
\sin^2 \theta_{13} \lsim 2.5 \times 10^{-2}$ and the original
electron antineutrinos have all been swapped for the more energetic $\mu$ and $\tau$
antineutrinos by the time they exit the supernova envelope, resulting in a harder incident
spectrum.
Thus, the initial $\neb$ spectrum would have to be softer for the inverted hierarchy than for the
normal hierarchy.

In performing our statistical analysis, we fix the solar neutrino oscillation parameters at 
$\Delta m^2_{21} = 3.7 \times 10^{-5}$ eV$^2$,
$\sin^2 2 \theta_{12} = 0.79$~\cite{lma} and the atmospheric neutrino scale 
$|\Delta m^2_{32}| = 3 \times 10^{-3}$ eV$^2$~\cite{fogli}
and fit the data taking $E_b$, $\tneb$ and $\sin^2 2 \theta_{13}$ as free parameters. The time 
structure of the SN 1987A signal (above threshold) is shown in Fig. 1, where it has been assumed
that the first events at the two detectors occurred simultaneously.
Most of the events are concentrated in the
first 2.75 seconds. We perform two sets of analyses to illustrate how sensitive 
any conclusion about the mass hierarchy is to the data sample chosen for analysis. For the first
set of analyses (which we label $t_{<13}$), we use all the data reported by the two experiments:
the 11 events observed by KII and the 8 events observed by IMB. 
For the second set (which we call $t_{<3}$), we only include events within the first 2.75 seconds: 
8 events at KII and 6
at IMB. 
All are assumed to be antineutrino events~\cite{sato}.
The procedure employed for our likelihood analysis is that of Ref.~\cite{raffelt} up to four minor
refinements listed below. (We refer the reader
to Ref.~\cite{raffelt} for the definition of the likelihood function and a 
description of the method).
\begin{enumerate}
\addtolength{\itemsep}{-2.5mm}
\item{We use an improved $\neb - p$ cross section which includes Coulomb, weak magnetism, recoil
and outer radiative corrections~\cite{beacom}.
It is approximately 8\% larger than the cross-section used in Ref.~\cite{raffelt}.}
\item{Instead of using the constant density approximation for the earth to calculate
$\bar{P}_{1e}$, we numerically integrate the evolution equations of neutrinos through a
realistic density profile of the earth~\cite{prem}.}
\item{We approximate the time integrated spectra of neutrinos with a Fermi-Dirac
distribution instead of a Boltzmann distribution.} 
\item{We use the detector efficiencies given in Ref.~\cite{burrows}.}
\end{enumerate}

To convey how these modifications (other than the different treatment of
earth matter effects) affect the $t_{<13}$ analysis, we show the 
1$\sigma$ and 2$\sigma$ allowed regions in $\eneb$ and $\tneb$ for the case of no neutrino mixing 
in Fig.~\ref{snno}. 
The results from separate analyses of KII and IMB data and an analysis of
the combined data are shown. This facilitates a direct comparison with other 
analyses~\cite{raffelt,valle}.  The dark-shaded region is the prediction
of supernova models from Eq.~(\ref{models}) while the light-shaded region is the result based on
Ref.~\cite{models2}. The overlap with the predicted $\tneb$ is significantly greater for
the $t_{<3}$ analysis, as can be seen in Fig.~\ref{snno1}. Also, KII and IMB data from the first 
three seconds are consistent at the 1$\sigma$ level. 

For the inverted hierarchy, we perform three-parameter $t_{<13}$ analyses 
for $\tau\equiv \tnx/\tneb = 1.7$ 
(corresponding to the middle of the ranges of Eq.~\ref{models}), 
$\tau=1.4$ (the lowest value of $\tau$ from Eq.~\ref{models}) and 
$\tau=1.25$ (corresponding 
 to a $\bar{\nu}_x$ spectrum that is softer~\cite{models2,models3,reddy} 
than traditionally obtained),
the results of which are shown in Figs.~\ref{sni1}-\ref{sni3}, respectively. The results of
the equivalent $t_{<3}$ analyses are shown in Figs.~\ref{sni11}-\ref{sni31}.
The figures show allowed regions at the 1$\sigma$ and 2$\sigma$ C.L. 
in the parameter space defined by
$E_b$, $\tneb$ and $\sin^2 2 \theta_{13}$. We do not consider $\sin^2 2 \theta_{13}>0.1$ since
this would violate the CHOOZ bound~\cite{chooz}.

\begin{table}[ht]
\begin{center}
\label{tab3par}
\begin{tabular}{|l|c|c|c|c|c|} 
\hline 
 & $\tau=\tnx/\tneb$ & $E_b$ ($10^{53}$ ergs)& $\tneb$ (MeV)& $\sin^2 2 \theta_{13}$ & $ln({\cal L}_{max})$\\ \hline
t $<$  13 sec &$1.25$ & 3.2 & 3.4 & 1.3 $\times {\rm 10}^{\rm -6}$ & -41.9\\ \hline
 &$1.4$ & 3.4 & 3.2 &  $1.3\times {\rm 10}^{\rm -6}$ & -41.6\\ \hline
 &$1.7$ & 4.5 & 2.6 & $4.0 \times {\rm 10}^{\rm -6}$ & -41.2 \\ \hline
 &$2.0$ & 6.3 & 2.0 & $1.6 \times {\rm 10}^{\rm -5}$ & -40.6 \\ \hline
t $<$  3 sec &$1.25$ & 2.0 & 3.6 & $1.6 \times {\rm 10}^{\rm -5}$ & -35.5\\ \hline
 &$1.4$ & 2.0 & 3.5 & $1.3 \times {\rm 10}^{\rm -6}$ & -35.4\\ \hline
 &$1.7$ & 2.4 & 2.9 & $1.3 \times {\rm 10}^{\rm -5}$ & -35.2 \\ \hline
 &$2.0$ & 3.0 & 2.5 & $3.2 \times {\rm 10}^{\rm -6}$ & -34.9
\end{tabular}
\end{center}
\caption{Best fit values for the binding energy $E_b$, electron
antineutrino temperature $\tneb$, and $\sin^2 2 \theta_{13}$
from three parameter fits to all the SN 1987A data and to data from the first three seconds, for the inverted neutrino mass hierarchy and several values of 
$\tau$. The corresponding logarithms of the likelihood
function are given in the last column.}
\end{table}

The best-fit points are displayed in Table~I, and as expected the $t_{<3}$ analysis gives 
higher best-fit $\tneb$ (since
the same is true for no oscillations), and the best-fit values of 
$\sin^2 2 \theta_{13}$ correspond 
to highly non-adiabatic transitions.
However, $\sin^2 2 \theta_{13}$ is not constrained even at the 1$\sigma$ level.
 Values of
$\tau$ greater than 1.7 lead to $\tneb$ shifted to smaller values than for $\tau=1.7$, but the
$\sin^2 2 \theta_{13}$ contour remains open at the 1$\sigma$ level. 
It can be seen qualitatively, that for $\sin^2 2 \theta_{13}\lsim 10^{-4}$, 
values of $\tneb$ from supernova codes are
less inconsistent with the data (see the lower panels of Fig.~\ref{sni1}-\ref{sni31}). 
But, for $\eneb$ in the range found in Ref.~\cite{models2} and $\tau \lsim 1.4$, 
SN 1987A data are consistent with theoretical predictions
even for $\sin^2 2 \theta_{13}$ as high as the CHOOZ bound. The results of Ref.~\cite{reddy} 
indicate that both the above conditions on $\eneb$ and $\tau$ can easily be met simultaneously. 
Thus, even at a qualitative level it can not be claimed that the inverted mass hierarchy
is disfavored for $\sin^2 2 \theta_{13}\gsim 10^{-4}$.

  For a given value of $\tau$, the
 range of $\tneb$ that is allowed for the normal hierarchy
is the same as that for the inverted hierarchy with $\sin^2 2 \theta_{13}=0$ since the
survival probabilities, Eqs.~(\ref{norprob}) and~(\ref{invprob}), are the same for both hierarchies 
in this limit.
The $\tneb$ for the normal hierarchy can be read-off from the lower panels of 
Figs.~\ref{sni1}-\ref{sni31} at $\sin^2 2 \theta_{13}=10^{-6}$.
(This interpretation must be made
with caution, since for the normal hierarchy, it is known a priori that the antineutrino 
spectra are independent of $\sin^2 2 \theta_{13}$, which would strictly mean that the confidence
regions should be determined for two parameters and not three). 

To make the fact that SN 1987A data did not probe the neutrino mass hierarchy even more
 transparent, we show the results of two-parameter fits in $\eneb$ and $\tneb$ 
for the normal hierarchy and the 
inverted hierarchy (with $\sin^2 2 \theta_{13}=0.01$) in Table~II ($t_{<13}$) and 
Table~III ($t_{<3}$).
We do not show figures of these allowed regions since these exist in the 
literature~\cite{raffelt,lunardini,valle,minakata} for the $t_{<13}$ case. 
As has been pointed out by previous authors, if 
the hierarchy is inverted and $\sin^2 2 \theta_{13}$ is large ({\it i.e.}, $P_H \approx 0$), 
lower values of $\tneb$ are required to fit the data than if the hierarchy is normal.
Since in all cases the likelihoods are comparable, the data itself
does not favor one neutrino mass hierarchy over the other. 
Any deductions about the hierarchy can only be based
on the disagreement of the data with supernova model predictions, which are in a state of 
change.

\begin{table}[t]
\begin{center}
\begin{tabular}{|l|c|c|c|} 
\hline 
 & $E_b$ ($10^{53}$ ergs) & $\tneb$ (MeV) & $ln({\cal L}_{max})$  \\ \hline
 no oscillations           & 3.2 & 3.6 & -42.0\\ \hline
 $\tau=1.25 \;,\; P_H = 0 $ & 3.1 & 2.9 & -42.0\\ \hline
 $\tau=1.25 \;,\;$ normal   & 3.2 & 3.4 & -41.9\\ \hline
 $\tau=1.4\  \;,\; P_H = 0 $ & 3.1 & 2.6 & -42.0\\ \hline
 $\tau=1.4\  \;,\;$ normal   & 3.4 & 3.2 & -41.6\\ \hline
 $\tau=1.7\  \;,\; P_H = 0 $ & 3.2 & 2.1 & -42.0\\ \hline
 $\tau=1.7\  \;,\;$ normal   & 4.2 & 2.7 & -41.2\\ \hline
 $\tau=2.0\  \;,\; P_H = 0 $ & 3.2 & 1.8 & -42.0\\ \hline 
 $\tau=2.0\  \;,\;$ normal   & 5.8 & 2.2 & -40.6
\end{tabular}
\label{tab2par}
\end{center}
\caption{Best fit values for $E_b$ and $\tneb$ from two-parameter fits to all the KII and IMB data. 
Results
are presented for the case in which no oscillations occur, the 
inverted hierarchy with $\sin^2 2 \theta_{13}=0.01$ ($P_H \approx 0$),
and for the normal hierarchy.}
\end{table}

\begin{table}[ht]
\begin{center}
\begin{tabular}{|l|c|c|c|} 
\hline 
 & $E_b$ ($10^{53}$ ergs) & $\tneb$ (MeV) & $ln({\cal L}_{max})$  \\ \hline
 no oscillations           & 1.8 & 4.0 & -35.6\\ \hline
 $\tau=1.25 \;,\; P_H = 0 $ & 1.8 & 3.2 & -35.6\\ \hline
 $\tau=1.25 \;,\;$ normal   & 1.9 & 3.7 & -35.5\\ \hline
 $\tau=1.4\  \;,\; P_H = 0 $ & 1.7 & 2.9 & -35.6\\ \hline
 $\tau=1.4\  \;,\;$ normal   & 2.0 & 3.5 & -35.4\\ \hline
 $\tau=1.7\  \;,\; P_H = 0 $ & 1.7 & 2.4 & -35.6\\ \hline
 $\tau=1.7\  \;,\;$ normal   & 2.4 & 3.0 & -35.2\\ \hline
 $\tau=2.0\  \;,\; P_H = 0 $ & 1.8 & 2.0 & -35.6\\ \hline 
 $\tau=2.0\  \;,\;$ normal   & 3.1 & 2.5 & -34.9
\end{tabular}
\label{tab2par1}
\end{center}
\caption{Same as Table~II but only data from the first 2.75 seconds are used.}
\end{table}

In addition to the above arguments, one must keep in mind that SN 1987A provided
us with very limited statistics with a somewhat uncertain time sequence.
Moreover, even with oscillations, the marginal overlap 
between the KII and IMB data should suggest that any conclusions can only be suggestive
at best. 

In summary, we have disputed the assertion that SN 1987A
provides a strong indication that
the inverted mass hierarchy is disfavored for 
$\sin^2 \theta_{13} \gsim 10^{-4}$~\cite{minakata}. The data provides
no substantial evidence that this is the case. 
Rather than telling us anything about the nature of the neutrino mass hierarchy, SN 1987A data
seem to indicate the need for more sophisticated supernova codes~\cite{models2} 
which could remove 
the (mass-hierarchy-independent) discrepancy with model predictions. For a future
galactic supernova, data from Super-Kamiokande and SNO would enable a precise determination of
 both $E_b$ and $\tneb$~\cite{inverting,minakata1} (unless $\sin^2 2 \theta_{13}\gsim 10^{-3}$ 
and the hierarchy is inverted~\cite{inverting})
and determine
 the sign of $\Delta m^2_{32}$ if $\sin^2 2 \theta_{13}\gsim 10^{-4}$~\cite{dighe,inverting},
independently of supernova models.

\vspace{0.5in}
{\it Acknowledgements}. 
We thank S. Reddy for numerous discussions and suggestions. 
This research was supported by the U.S.~DOE  
under Grants No.~DE-FG02-95ER40896 and No.~DE-FG02-91ER40676  
and by the WARF.

\newpage
\begin{figure}[t]
\centering\leavevmode
\mbox{\psfig{file=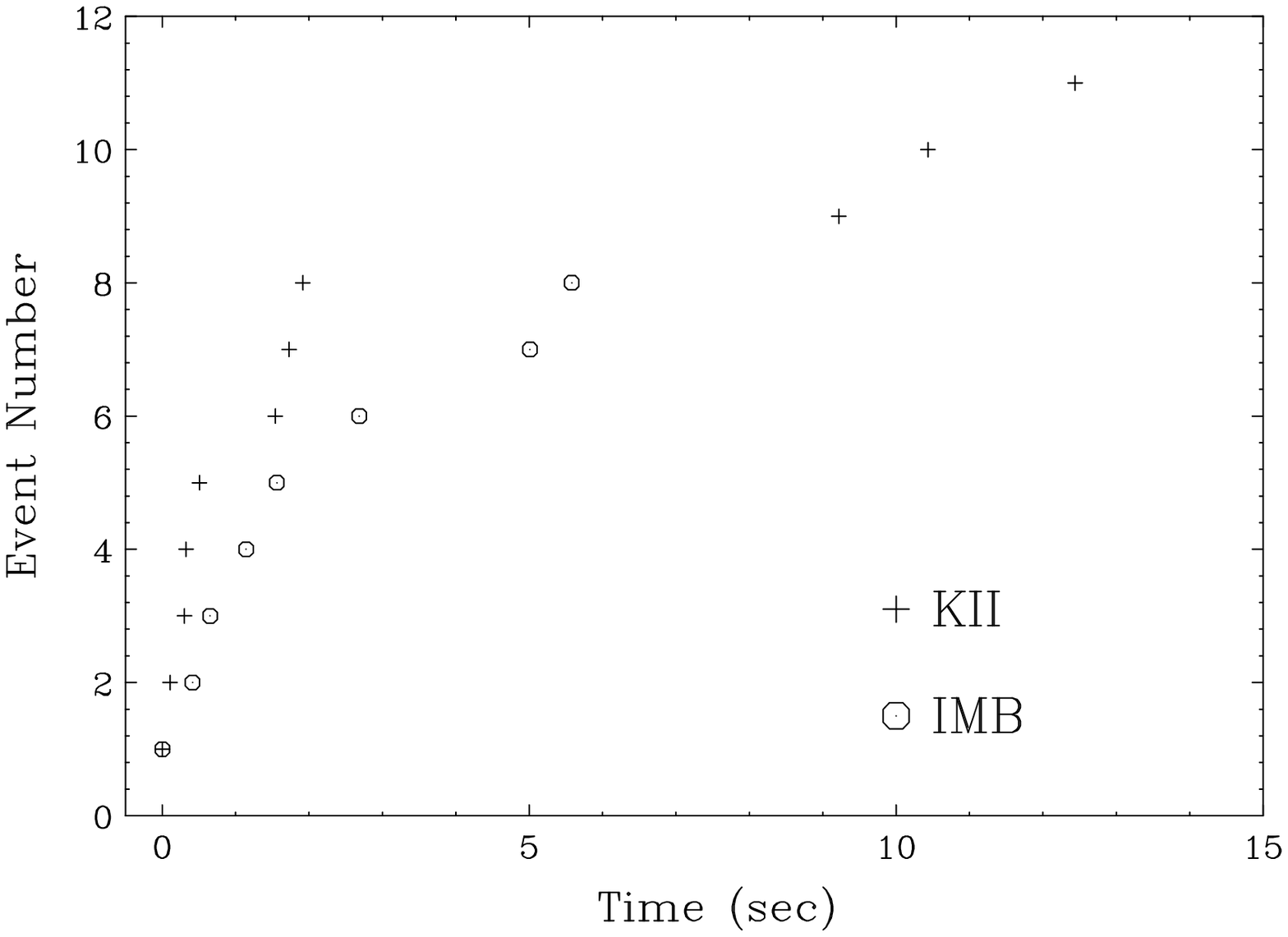,width=12cm,height=12cm}}
\medskip
\caption[]{The time structure of the SN 1987A signal. 14 of the 19 events occurred in the first 2.75
seconds.}
\label{time}
\end{figure}

\begin{figure}[t]
\centering\leavevmode
\mbox{\psfig{file=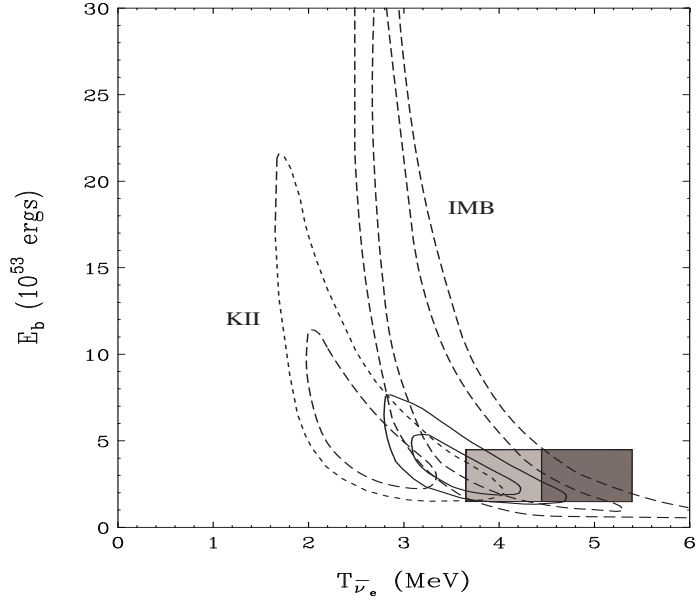,width=11cm,height=8cm}}
\medskip
\caption[]
{The 1$\sigma$ and 2$\sigma$ allowed regions for $E_b$ and $\tneb$ from a two-parameter fit 
to SN 1987A data in the case that no oscillations occur. The results from separate KII and IMB  
analyses (dashed), and a combined analysis (solid) of the data sets are shown. 
All data in the first
13 seconds are included.
The dark-shaded region is the range
of values from Eq.~(\ref{models}) and the light-shaded region 
is based on Ref.~\cite{models2}.}
\label{snno}
\end{figure}

\begin{figure}[t]
\centering\leavevmode
\mbox{\psfig{file=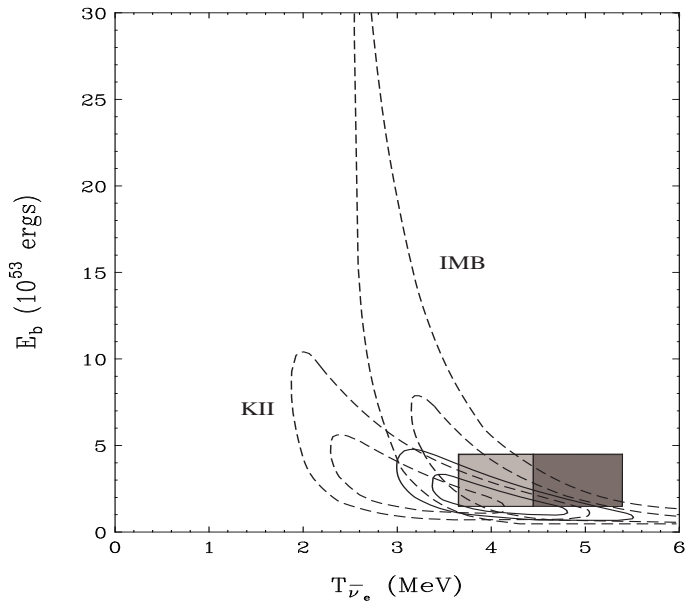,width=10cm,height=8cm}}
\medskip
\caption{Same as Fig.~\ref{snno} but only data from the first 2.75 seconds are included.}
\label{snno1}
\end{figure}

\begin{figure}[t]
\centering\leavevmode
\mbox{\psfig{file=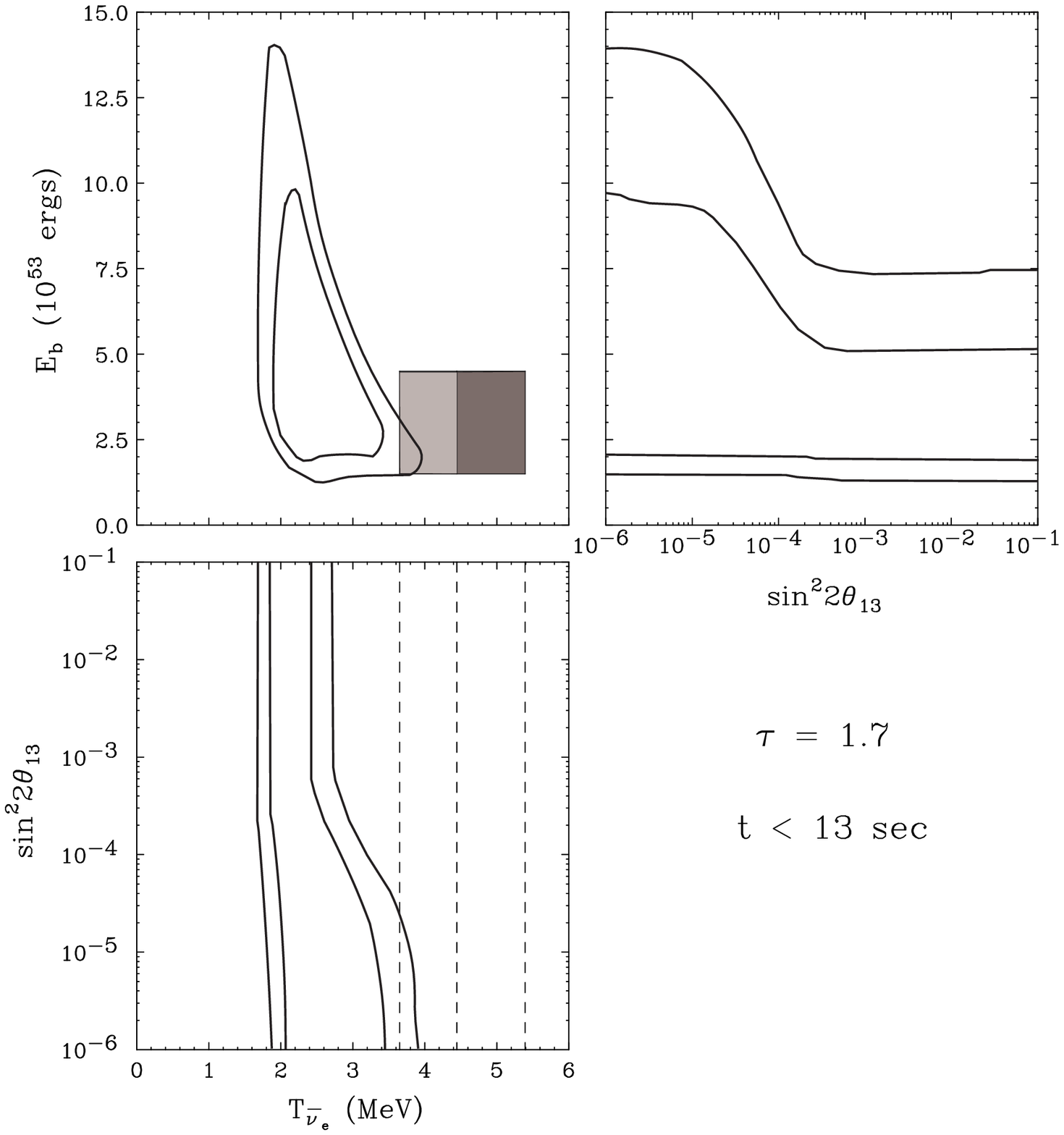,width=15cm,height=15cm}}
\medskip
\caption[]{1$\sigma$ and 2$\sigma$ allowed regions from a three-parameter $t_{<13}$ analysis of 
combined KII and IMB data for the inverted hierarchy. 
A ratio $\tau=T_{\bar{\nu}_x}/T_{\nu_e}=1.7$ is assumed. 
}
\label{sni1}
\end{figure}

\begin{figure}[t]
\centering\leavevmode
\mbox{\psfig{file=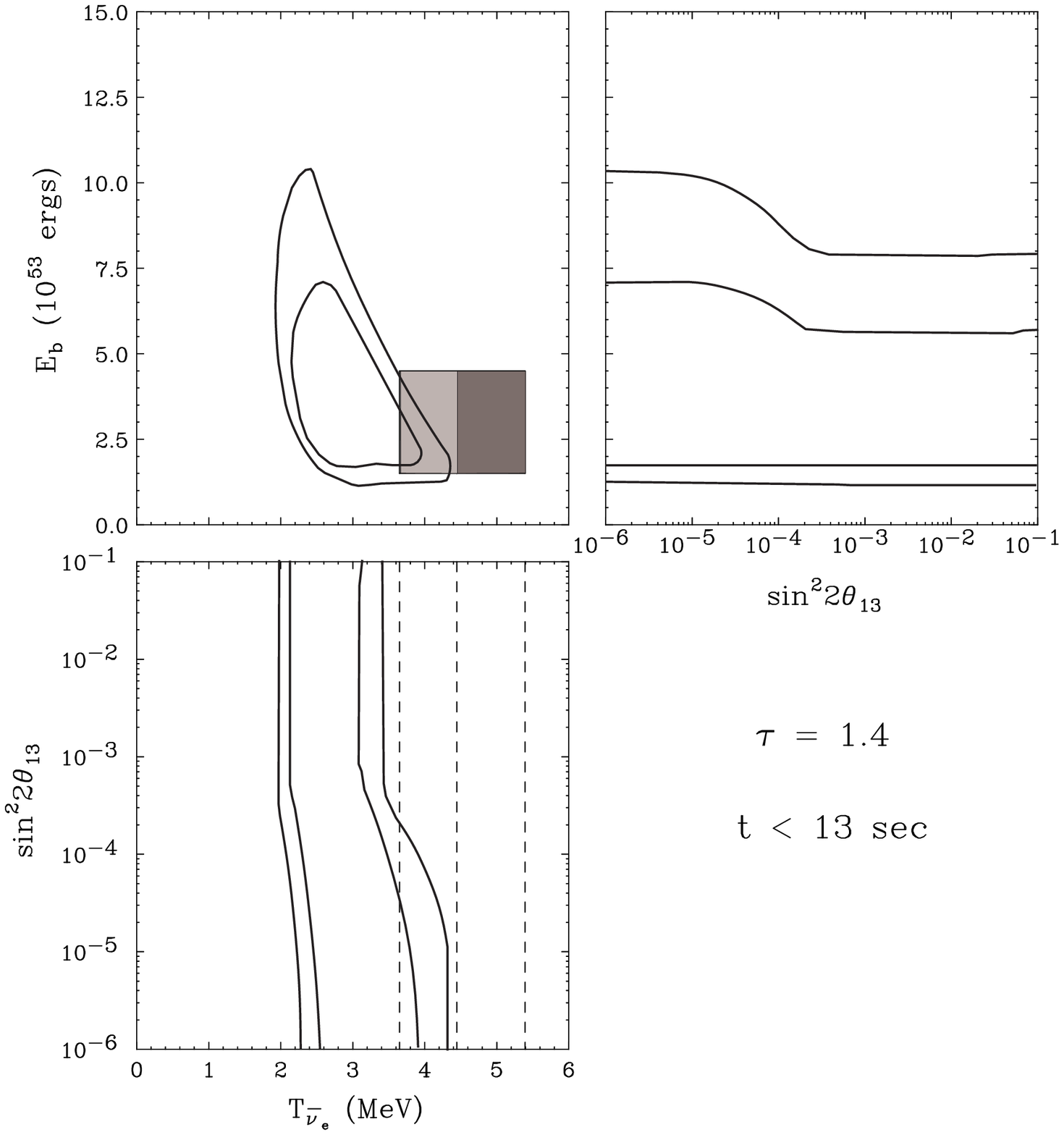,width=15cm,height=15cm}}
\medskip
\caption{Same as Fig.~\ref{sni1} but with $\tau=1.4$.}
\label{sni2}
\end{figure}

\begin{figure}[t]
\centering\leavevmode
\mbox{\psfig{file=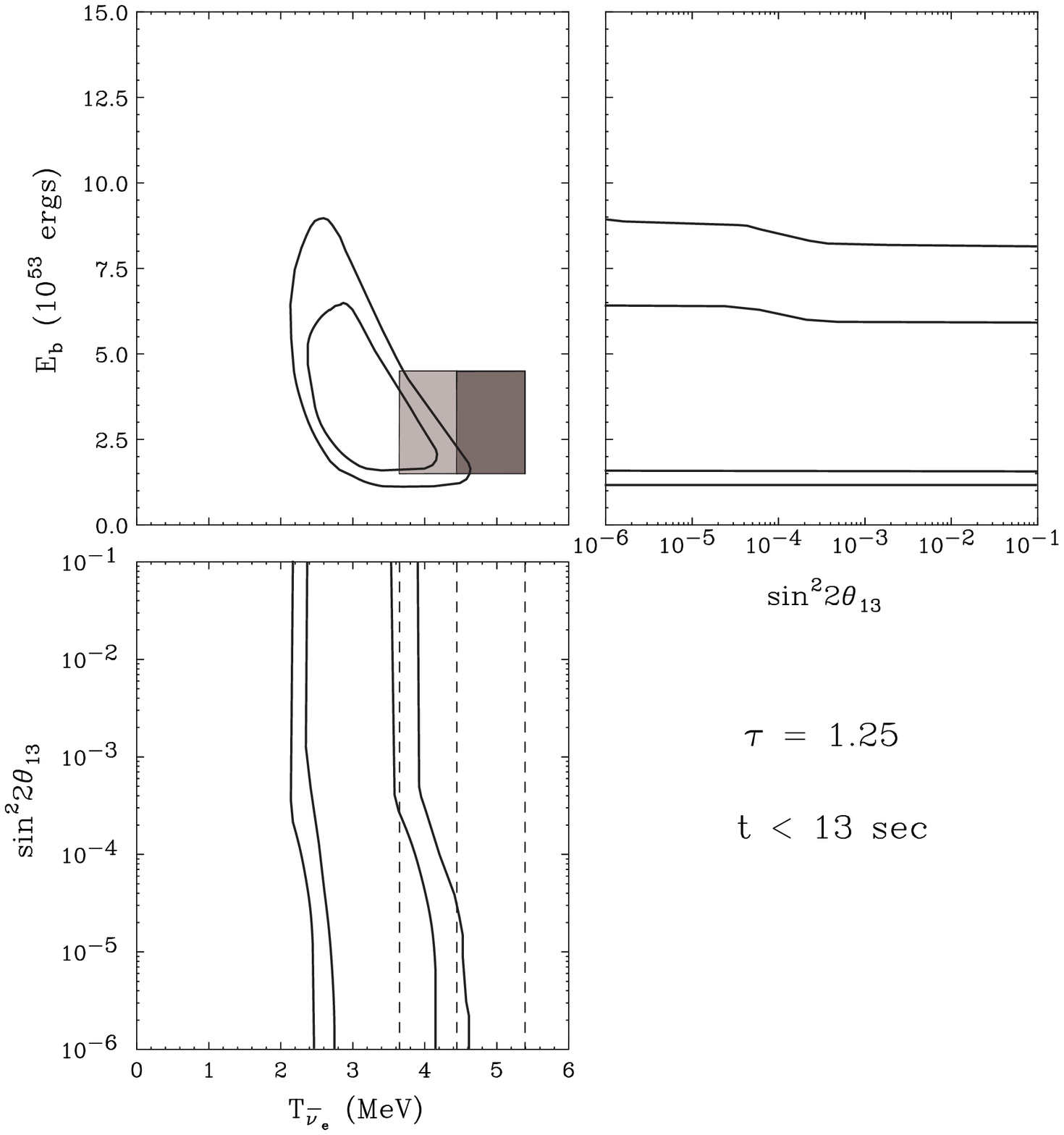,width=15cm,height=15cm}}
\medskip
\caption{Same as Fig.~\ref{sni1} but with $\tau=1.25$. }
\label{sni3}
\end{figure}

\begin{figure}[t]
\centering\leavevmode
\mbox{\psfig{file=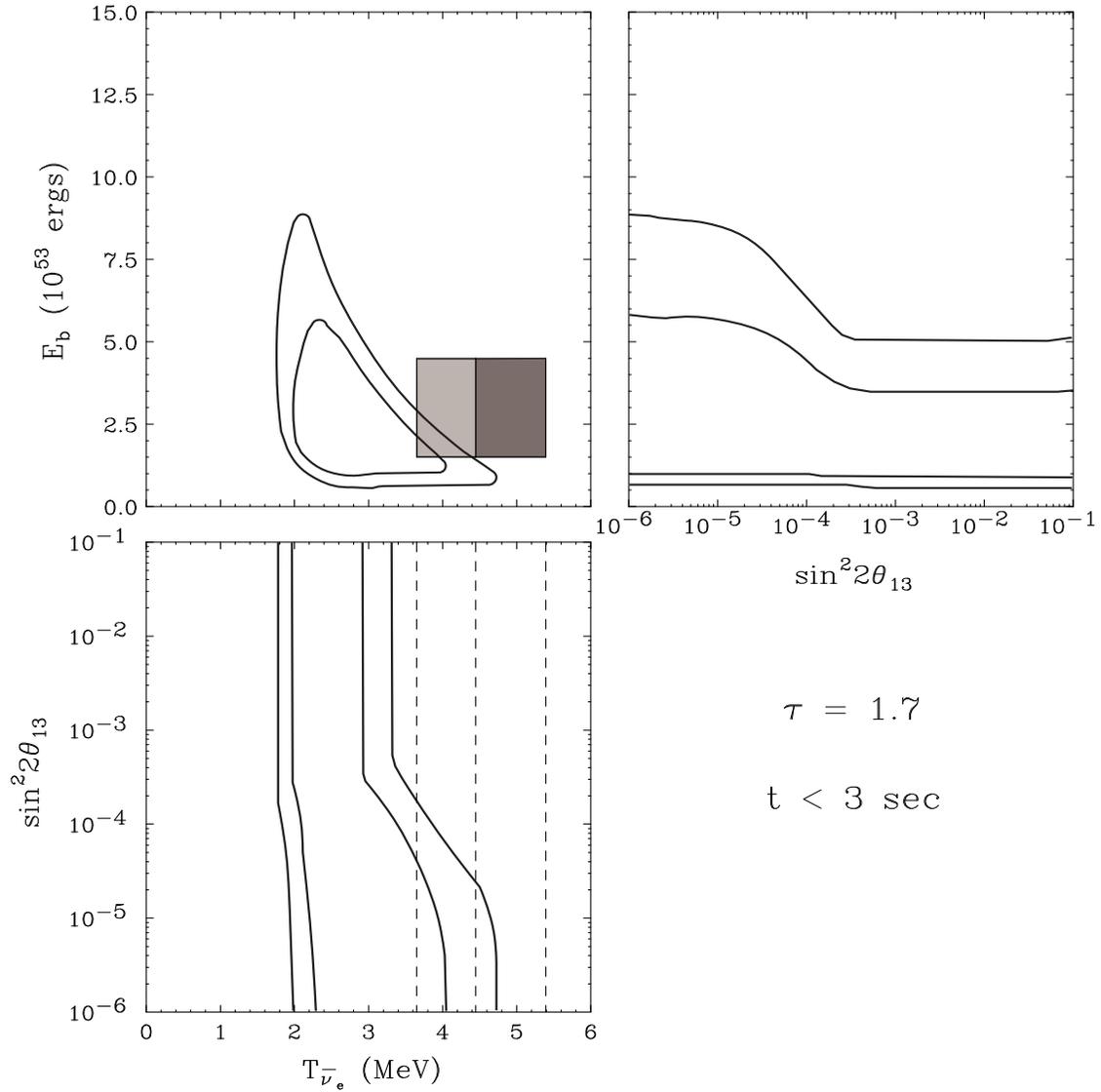,width=15cm,height=15cm}}
\medskip
\caption[]{1$\sigma$ and 2$\sigma$ allowed regions from a three-parameter $t_{<3}$ analysis of 
combined KII and IMB data for the inverted hierarchy. 
The ratio $\tau=T_{\bar{\nu}_x}/T_{\nu_e}$ is 1.7. 
}
\label{sni11}
\end{figure}

\begin{figure}[t]
\centering\leavevmode
\mbox{\psfig{file=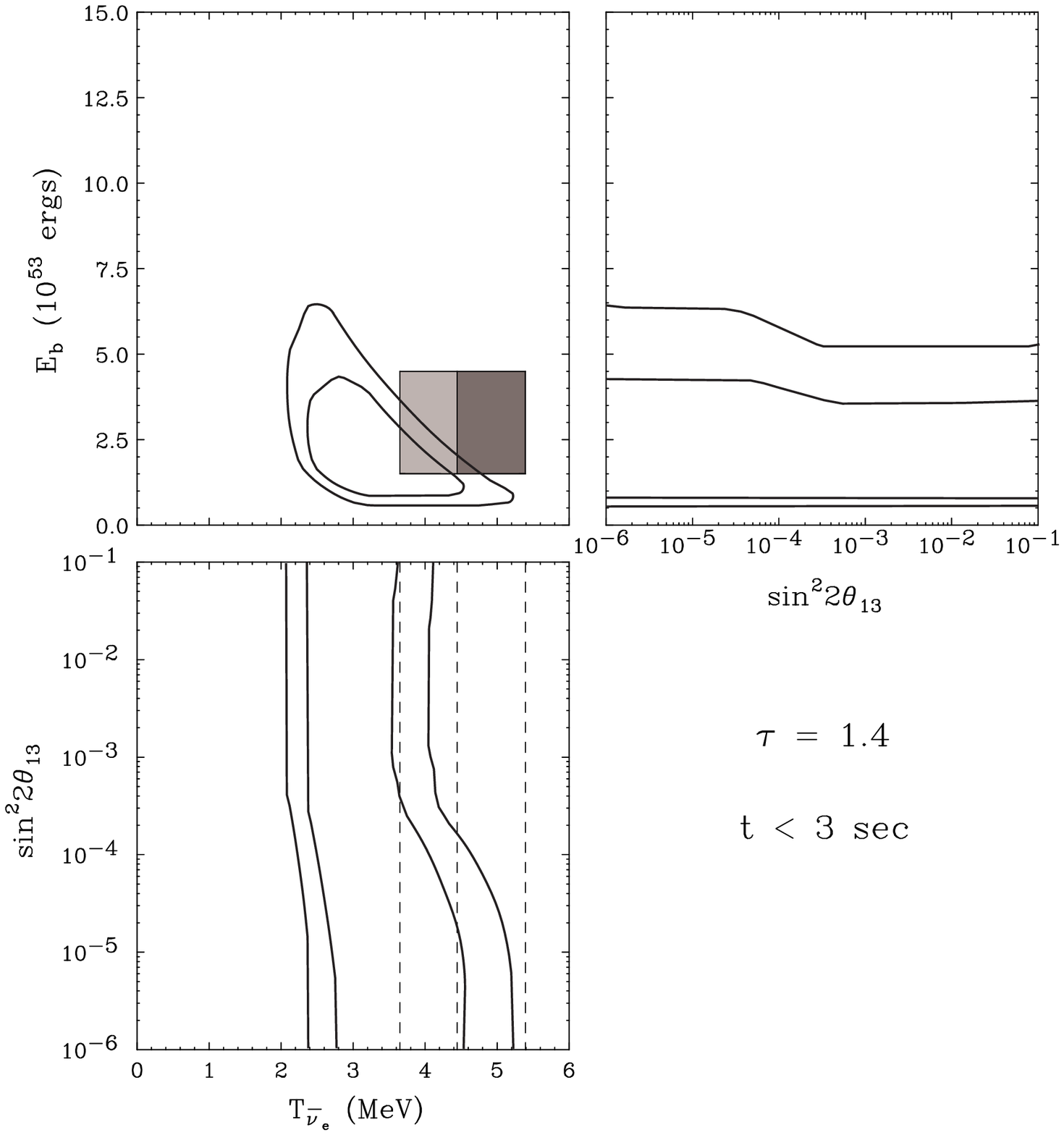,width=15cm,height=15cm}}
\medskip
\caption{Same as Fig.~\ref{sni11} but with $\tau=1.4$.}
\label{sni21}
\end{figure}

\begin{figure}[t]
\centering\leavevmode
\mbox{\psfig{file=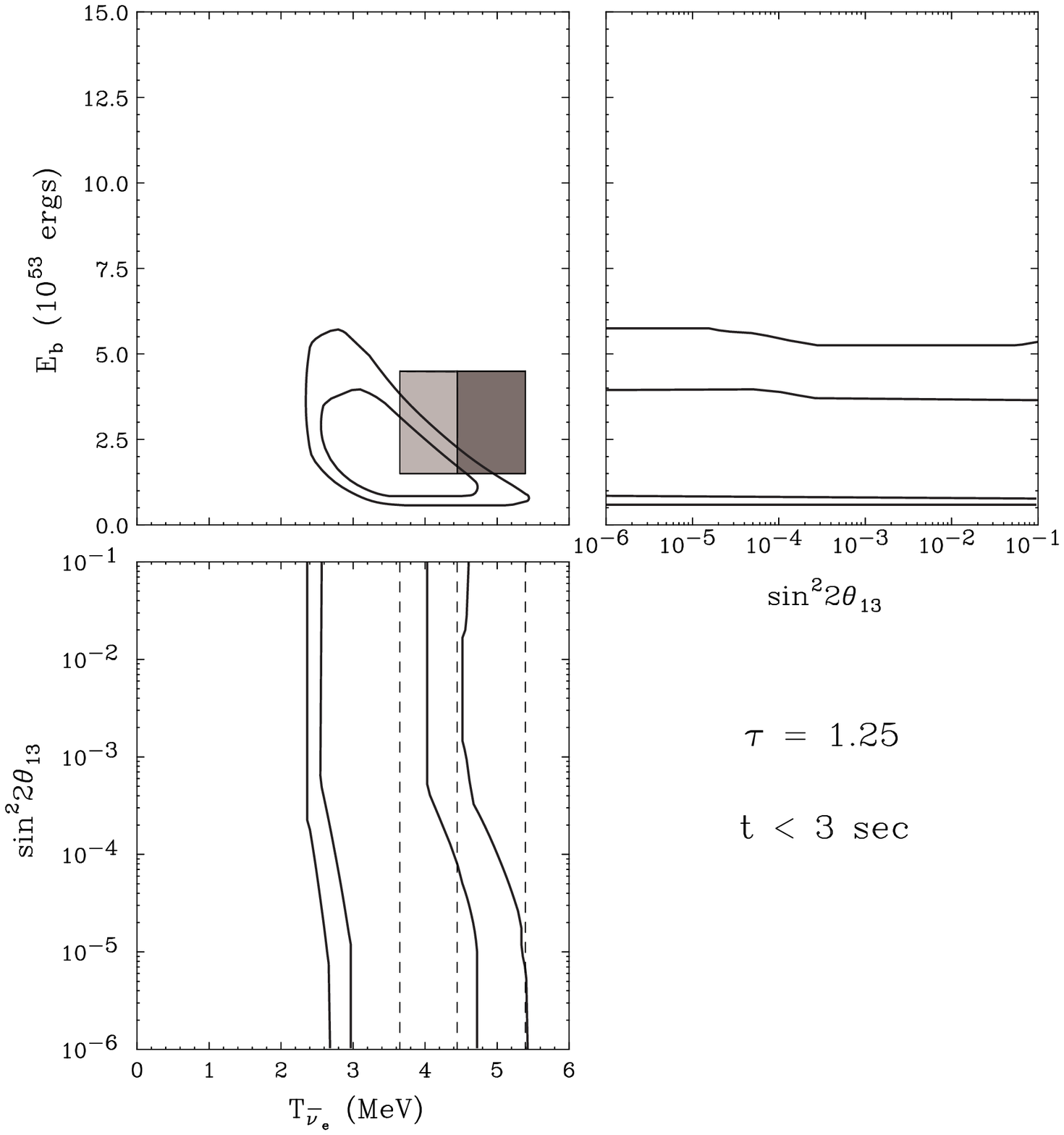,width=15cm,height=15cm}}
\medskip
\caption{Same as Fig.~\ref{sni11} but with $\tau=1.25$. }
\label{sni31}
\end{figure}

\end{document}